\documentclass{desyproc}

\begin{document}
%------------------------------------
\title{Light-Shining-Through-Walls with Lasers}

\author{{\slshape Friederike Januschek$^1$}\\[1ex]
$^1$Deutsches Elektronen-Synchrotron (DESY), Hamburg, Germany}

\contribID{Januschek\_Friederike}

\confID{300768}  % if the conference is on Indico uncomment this line
\desyproc{DESY-PROC-2014-03}
\acronym{Patras 2014} % if you want the Acronym in the page footer uncomment this line
\doi  % if there is an online version we will register DOIs

\maketitle

\begin{abstract}
Light-Shining-Through-Walls experiments are the search experiments for weakly interacting slim particles (WISPs) with the smallest model dependence. They have the advantage that not only the detection, but also the production of the WISPs takes place in the laboratory and can thus be controlled. Using lasers is the preferred option for most of the mass region and has led to the world's most stringent laboratory limits (ALPS~I) there. At CERN, OSQAR promises to surpass these and at DESY ALPS~II is currently set up, which is planning to probe the axion-like particle to photon coupling down to $|g_{a\gamma}|\gtrsim 2\cdot10^{-11}$ GeV$^{-1}$, which is in a region favored by many astrophysical hints.
\end{abstract}

\section{Motivation}
\label{sec:motiv}
WISPs are predicted by many extensions of the Standard Model. Most importantly, the axion provides the most elegant solution to the strong CP problem of quantum chromo dynamics~\cite{Peccei:1977hh, Weinberg:1977ma, Wilczek:1977pj}. In string theory inspired Standard Model extensions axion-like particles (ALPs) are expected~\cite{Ringwald:2012hr}. These theoretical motivations would already make it worthwhile to look for WISPs experimentally, but the cause for their existence gets much stronger by pointing out that axions and ALPs are also well-motivated (cold) dark matter candidates. 

\begin{wrapfigure}[20]{r}{0.55\textwidth}
\centerline{\includegraphics[width=0.55\textwidth]{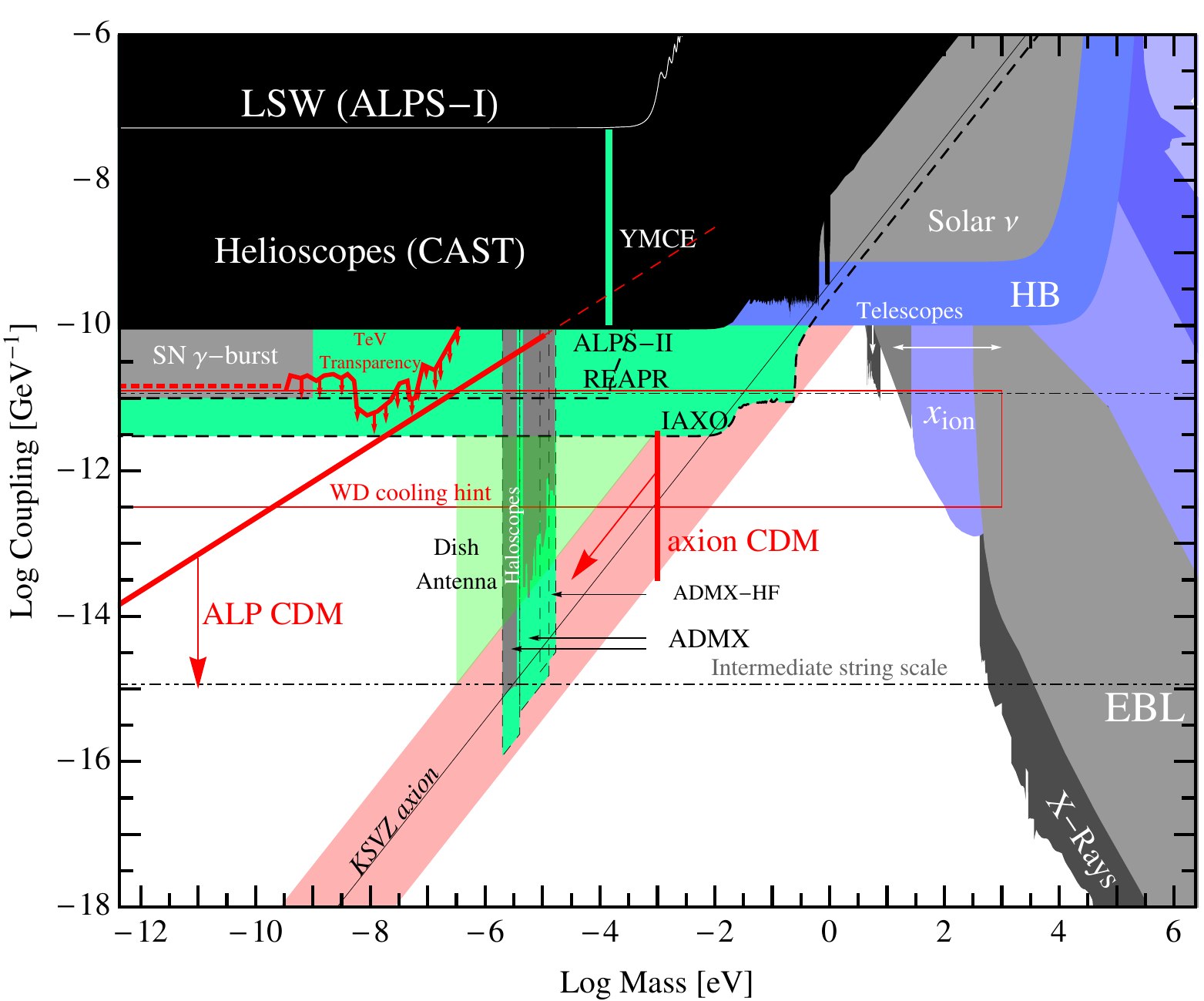}}
\caption{ALP parameter space with important hints and exclusion regions of past, present and future experiments~\cite{Essig:2013lka}.}
\label{fig:roi}
\end{wrapfigure}

Furthermore in the last years there have been several astrophysical phenomena, which could be explained by the existence of ALPs. Most prominently the reduced opacity of the universe for TeV photons compared to predictions~\cite{Horns:2012fx, Rubtsov:2014uga}: Extremely energetic $\gamma$-rays from cosmological sources should be attenuated by pair production through interaction with the extragalactic background light. Several studies have shown that the observed energy spectra do not fit this picture. This could be explained by the oscillation of photons into ALPs in ambient magnetic fields, which then travel unhinderedly due to their weak interaction and finally convert back to photons~\cite{De Angelis:2007dy, Meyer:2013pny}. For the relevant parameter region see Figure~\ref{fig:roi}.

Recent observations like the unexplained 3.55~keV emission line from galaxy clusters and Andromeda, which could also be attributed to ALP-related decaying dark matter (see e.g. ~\cite{Cicoli:2014bfa, Jaeckel:2014qea}) or signatures of unexplained energy losses in stars \cite{Ayala:2014pea,Viaux:2013lha}.

\pagebreak

\section{Experimental options}
\label{sec:exp}
There are different prominent options looking for WISPs when employing $g_{a\gamma}$:
\begin{itemize}
\item Haloscopes
\item Helioscopes and
\item Light-Shining-Through-Walls experiments.
\end{itemize}
Haloscopes like ADMX \cite{rosenberg} in Washington and WISPDMX \cite{lobanov} in Hamburg look for dark matter WISPs, typically using a resonant microwave cavity. This has the advantage of being very sensitive so that even the QCD axion is in reach, but only in a very narrow mass region (see Figure~\ref{fig:roi}). Also the results are theoretically dependent on cosmology and astrophysics. 

Helioscopes (like CAST \cite{Zioutas:2004hi}) look for WISPs produced in the sun, thus their results only depend on astrophysics, but they are also less sensitive. As they are very broadband, their limits are the constraining factor in a wide mass region (see Figure~\ref{fig:roi}). 

\begin{wrapfigure}{r}{0.6\textwidth}
\centerline{\includegraphics[width=0.5\textwidth]{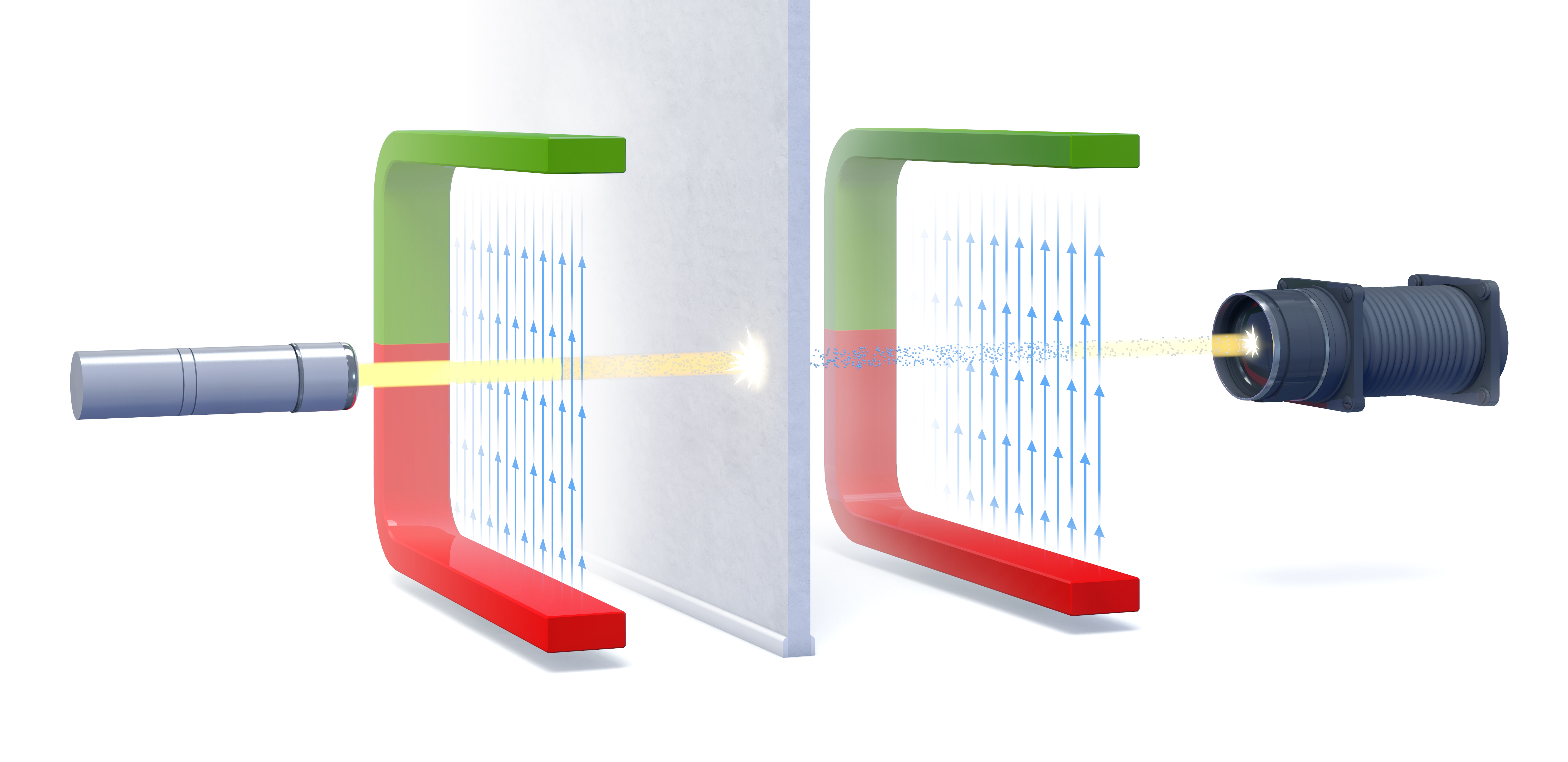}}
\caption{A diagram of a LSW setup~\cite{Bahre:2013ywa}: A strong light source is shone on an opaque wall. On the other side of the wall there is a sensitive light detector. In addition, magnets are required on both sides for ALP searches. }
\label{fig:lstw}
\end{wrapfigure}

In contrast to Haloscopes and Helioscopes, in Light-Shining-Through-Walls (LSW~\cite{Redondo:2010dp}, see Figure \ref{fig:lstw}) experiments the WISPs are produced (and detected) within the experiment. This has the distinct advantage that the theory dependence is kept at minimum and that the WISP production can be controlled and thus e.g. switched on and off.

In general the probability for a conversion between axions and photons $P_{\gamma\leftrightarrow a}$ due to the Primakoff effect 
 is described by 
\begin{equation}
P_{\gamma\leftrightarrow a}\propto(g_{a\gamma}BL)^2%\cdot|F|^2, 
\nonumber
\end{equation}
where $BL$ is the magnetic length and $g_{a\gamma}$ the coupling strength.
 
Thus the signal rate in a LSW experiment $\dot{N_{out}}$ is 
\begin{equation}
|\dot{N_{out}}|\propto \dot{N_{in}}\cdot (g_{a\gamma})^4 \cdot(BL)^4\cdot \epsilon_{det},
\nonumber
\end{equation} therefore depending on the number of incoming photons $ \dot{N_{in}}$, the coupling strength, the magnetic length and the detection efficiency $\epsilon_{det}$. 

LSW experiments can be done for all wavelengths, in principle. One example, using microwaves, is the CROWS experiment~\cite{Betz:2013dza}, having the advantage that operating low loss cavities is comparatively easy for microwaves. %Results are shown in Figure~\ref{fig:crows}. 
Microwaves do have the disadvantage though that their relatively low energy restricts the probed mass region. 

In contrast, LSW experiments can be (and have been~\cite{Battesti:2010dm, Inada:2013tx}) done using X-rays. In that case, high-mass regions can be investigated, but due to the impossibility of high-finesse cavities the number of photons and thus the sensitivity is rather limited.

Due to these reasons, LSW experiments in the visible/infrared with a laser as light source are the most common and several have been, are and will be performed (e.g.~\cite{Ehret:2010mh, Fouche:2008jk, Cameron:1993mr, Chou:2007zzc, Afanasev:2008jt,  pugnat, Bahre:2013ywa}), see also Figure~\ref{fig:alps-I} and the following Sections) since the suggestion of the principle~\cite{Okun:1982xi, Anselm:1986gz, VanBibber:1987rq}. 

\section{ALPS~I}
\label{sec:alps-I}

\begin{wrapfigure}{r}{0.45\textwidth}
\centerline{\includegraphics[trim=0.1cm 0.0cm 0.0cm 0.1cm, clip=true,width=0.5\textwidth]{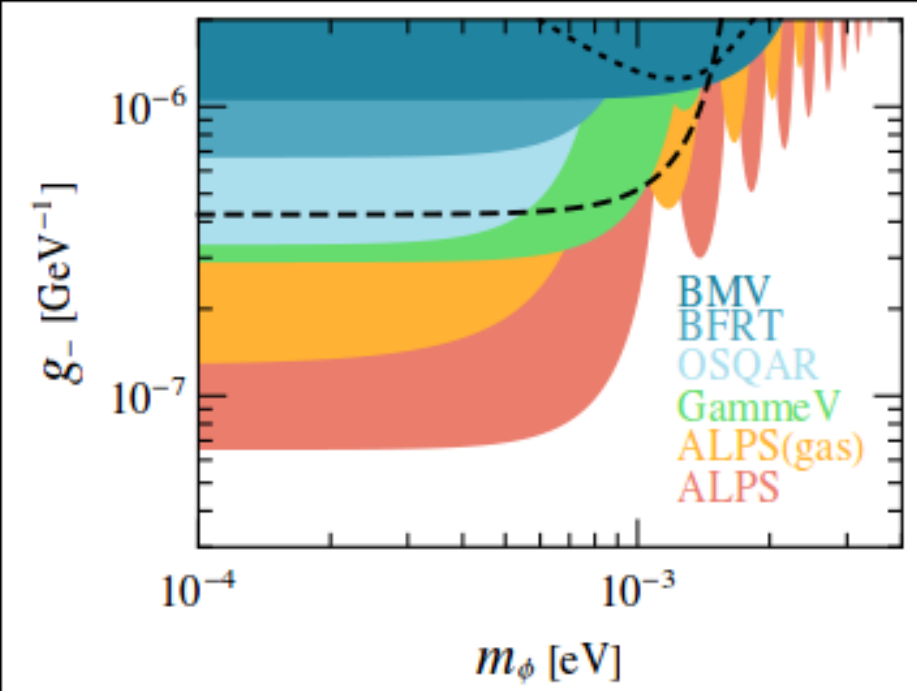}}
\caption{Exclusion limits (95\% C.L.) or scalar  axion-like-particles from the ALPS~I~\cite{Ehret:2010mh}, BFRT~\cite{Cameron:1993mr}, BMV~\cite{Fouche:2008jk}, GammeV~\cite{Chou:2007zzc} and OSQAR (\cite{Pugnat:2007nu}, see Section~\ref{sec:osqar}) LSW experiments as of 2010, taken from~\cite{Ehret:2010mh}.}
\label{fig:alps-I}
\end{wrapfigure}
The ALPS~I experiment at DESY~\cite{Ehret:2010mh} was an LSW experiment at DESY from 2007 to 2010. It used an old HERA magnet of 5~T and  about 1.2~kW of circulating laser power supplied by a laser power output of $\approx$~4.6W of 512~nm green light and an optical resonator with a power build-up of about 300. The result of the data-taking in 2010  were the worldwide best laboratory limits (see Figure~\ref{fig:alps-I}).

\section{OSQAR}
\label{sec:osqar}

\begin{figure}[b]
\center
{
\includegraphics[width=0.55\textwidth]{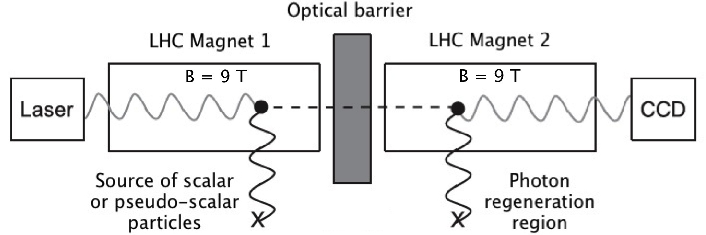}
}
\caption{Setup of the OSQAR experiment, taken from \cite{Pugnat:2013dha}.}
\label{fig:osqar1}
\end{figure}

OSQAR~\cite{pugnat, Pugnat:2007nu} is an experiment of the LSW type currently taking place at CERN. The setup can be seen in the left part of Figure~\ref{fig:osqar1}. At OSQAR an argon laser of 3-25~W and 488-514~nm is used in a 9 T transverse magnetic field of two LHC magnets over the unprecedented length of 2~$\times$~14.3~m. For detection a CCD camera is used. 

OSQAR was able to confirm the ALPS~I limits~\cite{Pugnat:2013dha}.

Further improvements are planned so that OSQAR could in the near future probe even deeper $g_{a\gamma}$ regions and look at regions previously not covered by LSW experiments.

\clearpage

\section{ALPS~II}
\label{sec:alps-II}
The ALPS~II experiment at DESY is the successor of ALPS~I with the aim of increasing the sensitivity reaching down to the main region of the astrophysical ALP hints. Therefore, ALPS~I is planned to be improved by three main factors: 
\begin{itemize}
\item an additional regeneration cavity, 
\item an increase in magnetic length through usage of more magnets and
\item a different detector system.
\end{itemize}
These changes, as described in the TDR~\cite{Bahre:2013ywa}, are supposed to lead to an increase in sensitivity to the coupling of a factor of about 3000 (see Table~\ref{tab:alps-II}).

\begin{wrapfigure}{r}{0.7\textwidth}
\centerline{\includegraphics[width=0.7\textwidth]{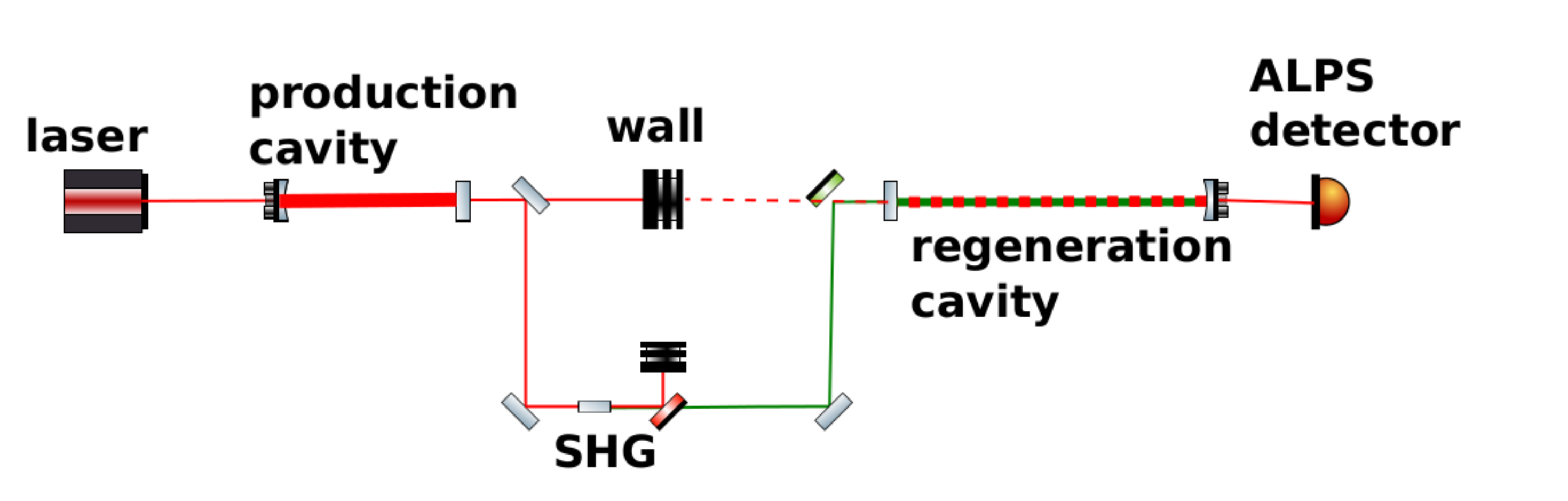}}
\caption{Setup of the ALPS~II experiment, adapted from~\cite{Bahre:2013ywa}.}
\label{fig:alps-II}
\end{wrapfigure}

A simple scheme of the ALPS~II setup is shown in Figure~\ref{fig:alps-II}. 
In addition to an increase of the power build-up of the production cavity in front of the wall, leading to an effective power of 150~kW, the regeneration cavity behind the wall is built to enhance the probability of WISP to photon reconversion probability by a factor of 40,000.

The most important factor with regards to a gain in sensitivity is the magnetic length increase. For that the usage of 20 HERA magnets (10 in front of and 10 behind the wall) is planned. As the HERA magnets are bent, the effective aperture -when using a long string of magnets- would be very small, leading to clipping losses and limiting the power build-up. Thus a method was devised, successfully tested and recently further improved to straighten the HERA magnets.

In contrast to ALPS~I, the light used in ALPS~II is near-infrared (1064~nm), which leads to the necessity of a different detector system as the quantum efficiency of a CCD would be very low (see~\cite{vonSeggern:2014dka}). Therefore, a Transition Edge Sensor (TES) detector system has been set up at the ALPS laboratory~\cite{Dreyling-Eschweiler:2014eya}. The system has been commissioned and calibrated with promising results, especially a low dark count rate, details see~\cite{dreyling}. 

The first stage of ALPS~II (ALPS~IIa, with two 10~m cavities and without magnets, thus searching only for hidden photons) is planned to start taking data in the end of 2014/beginning of 2015. The planned second-stage (ALPS~IIb, two 100~m cavities, but without magnets) had to be cut due to budget constraints. The final ALPS~II data-taking with the complete setup is scheduled for 2018.

\begin{table}
\centerline{
\begin{tabular}{|l|c|c|c|c|c|}
\hline
Parameter & Scaling & ALPS~I & ALPS~IIc & Sens. gain \\ [1pt] \hline
Effective laser power $P_{\rm laser}$ & $g_{a\gamma} \propto P_{\rm laser}^{-1/4}$ & 1\,kW & 150\,kW & 3.5\\[1pt] \hline
Rel. photon number flux $n_\gamma$ & $g_{a\gamma} \propto n_\gamma^{-1/4}$ & 1 (532\,nm) & 2 (1064\,nm) & 1.2\\[1pt] \hline
Power built up in RC $P_{\rm RC}$ & $g_{a\gamma} \propto P_{reg}^{-1/4}$ & 1 &  40,000 &  14\\[1pt] \hline
 $BL$ (before\& after the wall) & $g_{a\gamma} \propto (BL)^{-1}$ & 22\,Tm & 468\,Tm  & 21\\[1pt] \hline
Detector efficiency $QE$ & $g_{a\gamma} \propto QE^{-1/4}$ & 0.9 & 0.75 & 0.96\\[1pt] \hline
Detector noise $DC$ & $g_{a\gamma} \propto DC^{1/8}$ & 0.0018\,s$^{-1}$ & 0.000001\,s$^{-1}$ & 2.6\\[1pt] \hline
Combined improvements & &  &  & 3082\\[1pt] \hline
\end{tabular}
}
\caption{Sensitivity of ALPS~II compared to ALPS~I, taken from~\cite{Bahre:2013ywa}.}
\label{tab:alps-II}
\end{table}

\pagebreak
\section{Conclusion and outlook}
\label{sec:future}
LSW in the optical regime is an important tool for (relatively) model-independent searches for the well-motivated WISP particles. 
ALPS~I has set stringent laboratory-based limits and OSQAR will be able to improve them in the near future. 
ALPS~II, which is currently being set-up, will be able to reach the interesting region(s) with regards to the astrophysical hints. 

In addition, another LSW experiment with a laser, REAPR, has been proposed at Fermilab~\cite{Mueller:2009wt}, 
which is supposed to use a long string of Tevatron magnets and a production and a regeneration cavity similar to ALPS~II. For REAPR a heterodyne signal detection scheme has been suggested.

Looking at what would be possible to achieve with a LSW setup in the near future, the sensitivity could be increased by about a factor of 27 compared to ALPS~II (see Table~\ref{tab:alps-III}) through the use of better magnet (e.g. magnets as designed for the LHC upgrade) und further increasing the power build-up of the cavities. This sensitivity increase would then cover the complete region covered by the leading astrophysical  hints. Thus such an LSW experiment would either discover WISPs there or rule them out as an explanation for these phenomena.

\begin{table}
\centerline{
\begin{tabular}{|l|r|r|c|c|r|r|c|}
\hline
Exp. & Photon flux& $E_{\gamma}$&B&L&B$\cdot$L&PB&Sens.\\
& 1/s& eV&T&m&Tm&&rel.\\\hline%\hline
ALPS~I&$3.5\cdot 10^{21}$&$2.3$&$5.0$&$4.4$&$22$&$1$&$0.0003$\\\hline 
ALPS~IIc&$1\cdot 10^{24}$&$1.2$&$5.3$&$106$&$468$&$40,000$&$1$\\\hline 
Future&$3\cdot 10^{25}$&$1.2$&$13$&$400$&$5200$&$100,000$&$27$\\\hline 
%\hline
\end{tabular}
}
\caption{Comparison of ALPS~I, ALPS~II and a possible future experiment.}
\label{tab:alps-III}
\end{table}

 \section*{Acknowledgements}
 The author would like to thank the organisers of the motivating workshop and the members of the ALPS~II collaboration.

 \pagebreak
% ****************************************************************************
% BIBLIOGRAPHY AREA
% ****************************************************************************

\begin{footnotesize}

\end{footnotesize}

% ****************************************************************************
% END OF BIBLIOGRAPHY AREA
% ****************************************************************************

\end{document}